# Increased color conversion efficiency in hybrid light emitting diodes utilizing non-radiative energy transfer

S. Chanyawadee[1], P. G. Lagoudakis[1]*, R. T. Harley[1], M. D. B. Charlton[2], D. V. Talapin[3], and S. Lin[4]

[1]School of Physics and Astronomy, University of Southampton, Southampton, SO17 1BJ, UK
[2]School of Electronics and Computer Science, University of Southampton, Southampton, SO17 1BJ, UK
[3]Department of Chemistry, University of Chicago, Chicago, IL 60637, USA
[4]Luxtaltek Corporation, Chunan, Miaoli 350, Taiwan, R.O.C.

There have been numerous efforts to increase the efficiency of solid-state lighting, light-emitting diodes (LEDs) and displays during the last decades.[1-3] As the technologies for fabricating GaN-based LEDs and for synthesizing semiconductor colloidal nanocrystals (NQDs) mature, hybrid NQD-GaN LEDs are becoming promising candidates for highly efficient multi-color lighting. The high quantum yield and photostability of colloidal NQDs offer the possibility for flexible, low cost, large area, and simply-processed optoelectronic devices, while their emission color can be tuned from the visible to the near infrared range by either changing their size or chemical composition.[4] Also the epitaxial growth of GaN has now reached the stage where GaN-based LEDs have an internal quantum efficiency of 80%.[5] Although their external quantum efficiency is inevitably limited by total internal reflection due to the high refractive index contrast with air, several approaches to improve the out-coupling efficiency have been realised implementing smart photonic crystal and waveguide designs.[6,7] Color conversion LEDs consisting of colloidal NQD emitters pumped by GaN-based LEDs overcome the drawback of NQDs, i.e. low carrier transfer.[8] A thin NQD layer deposited on LED surface absorbs the high energy photons that are electrically generated in the LED and subsequently reemits lower energy photons. As a result, there is no charge transfer among colloidal NQDs involved in this color conversion process. However, the efficiency of radiative energy transfer is relatively low, <10%, due to several energy loss steps in the transfer process, i.e. waveguided leaky mode losses, light scattering from the NQDs and the reduced efficiency of emission in the blue, absorption and reemission at longer wavelengths from the NQD layers.

A number of approaches to improve the efficiency of electrical carrier injection into NQD emitters are of interest. A blend of semiconducting polymers and NQDs, for instance, employs the comparatively better carrier transfer properties of the semiconducting polymers to inject carriers into NQDs.[1,2] Integration of NQD monolayers directly into a semiconductor p-n junction diode possessing high carrier mobility was alternatively demonstrated.[9] Nevertheless, the performance of these devices is not comparable with conventional epitaxial semiconductor LEDs. The problems of charge imbalance at the emissive layer associated with different charge transport efficiency of polymer transport layers and the difficulties in integrating NQDs into a semiconductor p-n junction diode remain challenging. A new scheme of pumping NQDs by using nonradiative energy transfer has been demonstrated both theoretically[10,11] and experimentally.[12,13]

Nonradiative energy transfer is a dipole-dipole interaction between donors and acceptors where the excitation energy of a donor is transferred to an acceptor.[14] The nonradiative energy transfer rate ($k_{ET}$) scales linearly with spectral overlap and is proportional to $R^{-C}$, where $R$ is the donor-acceptor separation distance and $C$ is a constant. For example, $C$ = 2 or 6 describes layer to layer or point dipole to dipole energy transfer respectively.[15,16] In either case, to increase the energy transfer rate and consequently color conversion efficiency, the donor-acceptor separation distance has to be minimized. Hence, for colloidal NQD/multiple quantum well (QW) LED configuration, the QW barrier and the top contact layer has to be as thin as possible, while remaining thick enough in order to minimize surface recombination of injected carriers and to allow for uniform spreading of the injected carriers over the active layers.

In this report, we propose a novel method for utilizing nonradiative energy transfer in color conversion lighting by depositing bright colloidal NQDs on surface-patterned GaN-based LEDs. Elliptical cross-section holes are fabricated on the LED surface that penetrate through the active QW layers. We refer to this structure as a *deep etched* LED. Unlike conventional color conversion LEDs, this approaches brings colloidal NQDs (acceptors) into the close vicinity of the active layers (donors). The NQD absorption is tuned to strongly overlap with the LED emission and thus the two main requirements for efficient nonradiative energy transfer are satisfied. Electrically injected carriers in the active layers of the deep etched LED can be efficiently transferred to the adjacent NQDs by means of nonradiative energy transfer. A control structure, where the patterning of the surface does not penetrate the active layers is also fabricated and referred as a *shallow etched* LED. The donor-acceptor separation distance in the hybrid NQD/shallow etched LED is large and so nonradiative energy transfer does not occur. By comparing the hybrid NQD/deep etched LED with the hybrid NQD/shallow etched LED, the hybrid NQD/deep etched LED exhibits a 2-fold enhancement of NQD emission under the same emission intensity of blue photons from the QWs.

A schematic diagram of the hybrid NQD/deep etched LEDs is shown in Fig. 1a. A 2 μm thick undoped GaN buffer layer is grown on a sapphire substrate followed by a 2 μm thick *n*-doped GaN. The active layer consists of 5 InGaN/GaN QWs and a 50 nm thick *p*-doped AlGaN layer is grown on top. The device is terminated with a 0.2 μm thick *p*-doped GaN layer. An array of 450 nm deep slots of elliptical cross-sections is etched through the top layers penetrating through the multiple QWs (see experimetal reference). Figure 1b presents a schematic diagram of the hybrid NQD/shallow etched LED where an array of 150 nm deep slots are fabricated on the LED. The array patterns of both deep etched and shallow etched LEDs are identical. The inset of Fig. 1c shows a scanning electron microscope (SEM) image of the shallow etched LED surface. The array forms a photonic structure designed to improve the emission from the LED; its details are not relevant to our experiment as the same design has been used for both deep and shallow etched LEDs. The photoluminescence peak of the LED is at 460 nm at room temperature as shown in Fig. 1c. Highly efficient CdSe/CdS core/shell NQDs capped with hexadecylamine, tri-octylphosphine oxide (TOPO), and tri-n-octylphosphine (TOP) are used as color converters in this study.[17] The absorption and fluorescence spectra of the colloidal NQDs are illustrated in Fig. 1c.

Electroluminescence (EL) of the hybrid NQD/deep etched LED devices are measured perpendicularly to the surface and compared to that of the control devices (shallow etched LED). The EL intensity of the bare QW scales linearly with injection current and saturates at high injection current (not shown here). The linear increase of EL intensity implies that injected carriers form excitons in the QWs. The etching process partly removes active layers and inevitably introduces surface states to the side walls of the patterned structure. Consequently, the deep etched LED emission is reduced in comparison to the control device for the same injection current. In Fig. 2a, EL of the deep etched LED at 7 mA (pink dotted line) is comparable with that of the shallow etched LED at 3.8 mA (blue dotted line). The NQD emission pumped by the deep etched LED is compared with NQD emission pumped by the shallow etched LED to investigate the effect of nonradiative energy transfer on color conversion efficiency. For the same EL of the bare QW emission (460 nm), NQD fluorescence pumped by the deep etched LED (pink solid line) is significantly higher than NQD emission pumped by the shallow etched LED (blue solid line). Color-conversion efficiency ($\eta_c$) is defined as the ratio of the NQD emission intensity in the hybrid structure ($I_{NQD}^{H}$) to the bare QW emission intensity ($I_{QW}$), $\eta_C = I_{NQD}^{H} / I_{QW}$.[13] The average color-conversion efficiency of the hybrid NQD/deep etched LED in the linear regime is 20%, i.e. 43% increase from that of the hybrid NQD/shallow etched LED.

We propose that the enhancement of NQD emission is due to nonradiative energy transfer. Charge transfer through the insulating ligands of NQDs is inhibited due to the large band-offset between the carriers in the nanocrystals and surface molecules as well as between the nanocrystals and the QWs.[18] Effective nonradiative energy transfer efficiency ($\eta_{ET}^{*}$) is estimated from $\eta_{ET}^{*} = 1 - I_{QW}^{H} / I_{QW}$, where $I_{QW}^{H}$ is the emission intensity of QW donors in the presence of NQD acceptors. Here, $\eta_{ET}^{*}$ is averaged over the active area including both carriers that undergo nonradiative energy transfer and carriers that do not. At low injection current, $\eta_{ET}^{*}$ decreases with increasing injection current and remains constant for injection current above ~4 mA. It is conceivable that with increasing the injection current heating and Coulomb screening leads to exciton dissociation into free electron-hole pairs, hence the decrease of $\eta_{ET}^{*}$.[11] Figure 2b shows average enhancement of NQD emission, i.e. the ratio of NQD emission pumped by the deep etched LED to that pumped by the shallow etched LED. Evidently, the NQD fluorescence is enhanced through the whole range of injection currents and slightly decreases with increasing injection current. Insets of Fig. 2a show top-view image of the hybrid NQD/shallow etched LED (left panel) and of the hybrid NQD/deep etched LED (right panel). The corresponding images with a filter to cut the QW EL are shown in the inset of Fig. 2b.

To verify the existence of nonradiative energy transfer in the hybrid NQD/deep etched LED, we investigate the transient carrier dynamics in both the donor (QWs) and acceptor (NQDs) sites. The devices are excited at 400 nm with 100-fs pulse width and 250 kHz repetition rate. The photoluminescence is coupled into a streak camera with 300 ps resolution. Time-resolved data of the shallow etched LED and deep etched LED are illustrated in Fig. 3a and 3b respectively. The inset of Fig. 3a shows a streak image and the energy window used to extract the QW dynamics. The carrier dynamics of the unetched QWs with (green solid line) and without (red solid line) deposited NQDs are virtually identical. These results suggest that in the NQD/shallow etched LED where donor-acceptor distance is large, i.e. 100 nm, nonradiative energy transfer does not occur. Hence, carriers in deposited NQDs are only generated from conventional radiative energy transfer. In contrast, carrier dynamics of the etched QWs change after depositing NQDs as shown in Fig. 3b. The QW photoluminescence with deposited NQDs decays slightly faster over the first 4 ns and slower at later time compared to the bare QW. We interpret the faster decay as an effect of nonradiative energy transfer that introduces an additional decay channel for carriers in the QW. The slower decay at later times is believed to be an effect of surface passivation of the etched QWs from the organic ligands of the deposited NQDs.[19]

Since not all electron-hole pairs in the QWs undergo nonradiative energy transfer, we categorize the electron-hole pairs into two groups, i.e. the ones that undergo energy transfer and the ones that do not. As nonradiative energy transfer rate decreases dramatically with increasing donor-acceptor distance, electron-hole pairs close to the slots filled with NQDs (~10 nm distance) can undergo nonradiative energy transfer but the electron-hole pairs further away do not. The photoluminescence decay of the QW can be approximated with an exponential for the first 4 ns as shown in the inset of Fig. 3b. To extract the actual nonradiative energy transfer rate ($k_{ET}$) and the percentage of electron-hole pairs that undergo nonradiative energy transfer, we approximate and fit the photoluminescence decay of the QWs in the presence of the NQDs with the following equation,

$$I_{QW}^{H}(t) = A \cdot e^{-k_{QW}t} + B \cdot e^{-(k_{QW}+k_{ET})t}, \tag{1}$$

where $A$ and $B$ are the fraction of electron-hole pairs that undergo and do not undergo nonradiative energy transfer respectively. $k_{QW}$ is the total decay rate of the etched QWs taking into account the effect of surface passivation. From the fitting, 18% of generated electron-hole pairs experience nonradiative energy transfer with efficiency ($\eta_{ET} = \dfrac{k_{ET}}{k_{ET} + k_{QW}}$) of 82%. The energy transfer efficiency is high because the donor-acceptor distance is limited only by the short ligand molecules on the surface of the NQDs, i.e. ~2 nm. In previous studies of planar heterointerfaces efficiencies as high as 65% were reported where a 2.5 nm thick capping layer was present to separate a single QW from NQDs.[20] The absence of any barrier in this study allows for even higher nonradiative energy transfer efficiency to be observed.

Using the SEM images of our sample, as shown in Fig. 1c, to obtain the total perimeter of the slots we can estimate that nonradiative transfer occurs from only 2.3% of the QW material, assuming the process occurs from a 10 nm thick layer in the QWs. Thus under uniform excitation of our structure only ~3% of all the generated excitons in the active layer would undergo energy transfer in a static picture, which is approximately a third of the value we obtain from fitting the photoluminescence decay. However, exciton diffusion from regions further away from the holes towards the vicinity of the holes should increase the percentage of excitons undergoing nonradiative energy transfer. To investigate this possibility, we perform a 2-dimensional Monte Carlo simulation to calculate the percentage of excitons that undergo nonradiative energy transfer by taking into consideration exciton diffusion in the active layer of the deep etched LED. In our model, the motion of excitons in the active region is dominated by their thermal energy and their momentum scattering probability; scattering with impurities or phonons is described by scattering time, $\tau_s$, while the probability of radiative and nonradiative decay is given by $k_{QW}$ and $k_{ET}$, which we obtained from time-resolved measurements above. The momentum scattering time of carriers in InGaN QWs we estimate from $\tau_s = m^* \mu / e$, where $m^*$ is the effective mass, and μ is the carrier mobility. Although the carrier mobility in InGaN/GaN QWs varies widely in the literature[21,22] for a momentum scattering time in the range of 0.01 ps to 1 ps we obtain that ~18% of excitons undergo nonradiative energy transfer. This result is in reasonable agreement with the percentage obtained from fitting the photoluminescence decay with Eq. (1) and therefore lends support to the interpretation.

Observation of nonradiative energy transfer from the donor site does not provide conclusive evidence that the energy is transferred to the desired acceptor sites, namely the NQDs in our case. Thus it remains imperative to prove the effect of nonradiative transfer on the transient NQD dynamics. Figure 4a and 4b show the fluorescence decay of NQD deposited on etched and unetched QW LEDs respectively. It is apparent that the fluorescence of the former is higher due to nonradiative energy transfer. To exclude any quantitative effect that may originate from the larger number of NQDs in the patterned device we further investigate the effect of nonradiative energy transfer on the NQD fluorescence decay. The normalized data of Fig. 4a is subtracted by that of the normalized data of Fig. 4b. Figure 4c show that the energy transfer creates additional carriers in the deposited NQDs at early time as shown by the blue area (positive counts). A 20-nm window centred at the NQD fluorescence peak, as depicted with dashed lines in Fig. 4c, is used to extract the average carrier dynamic of the deposited NQDs shown in Fig. 4d. The carrier dynamic of the deposited NQDs including nonradiative energy transfer from the QWs can be described by the following equation,[23]

$$I_{NQD}^{H}(t) \propto \frac{k_{ET}}{k_{NQD} - k_{QW} - k_{ET}} \left( e^{-(k_{QW} + k_{ET})t} - e^{-k_{NQD}t} \right), \qquad (2)$$

where $k_{NQD}$ is the NQD fluorescence decay rate. By using the parameters obtained form the measurement of the time-resolved QW dynamics, the fitted carrier dynamic due to energy transfer from the QWs to the NQDs, as shown by a dotted line in Fig. 4d, is in good agreement with the experiment (solid red line), unequivocally demonstrating nonradiative energy transfer from the deep etched LED to the NQDs.

We propose a novel color conversion LED consisting of a surface-patterned blue LED and colloidal NQDs that promotes efficient nonradiative energy transfer from a higher energy light source to lower energy emission fluorophores. A 2-fold enhancement of the color conversion is demonstrated and attributed to nonradiative energy transfer by studying the transient dynamics both at donor and acceptor sites. This new design introduces a high performance color conversion LED with applications in solid-state lighting, displays and lasers from fluorophores of low carrier mobility.

*Experimental*

Nano-imprinting lithography is used to fabricate patterns onto LEDs. A 50 nm thick $SiO_2$ layer is initially deposited onto the LEDs by plasma enhanced chemical vapor deposition (PECVD) followed by a spin-coated polymer layer. A mold is brought into contact with the polymer layer under certain pressure. By applying heat above the glass transition temperature of the polymer, the pattern is transferred from the mold to the polymer layer. The LED samples and the mold are then cooled down to room temperature to release the mold. A reactive ion etching (RIE) with $CF_4$ plasma is used to remove the residual polymer layer and transfer the pattern onto $SiO_2$.

The patterns of different depths are eventually transferred to the LEDs by using inductively coupled plasma reactive ion etching (ICP-RIE). $Cl_2$ and Ar etching gases are introduced into the reactor chamber through independent electronic mass flow controllers (MFCs) that can control the flow rate with an accuracy of about 1 sccm. An automatic pressure controller (APC) is placed near the exhaust end of the chamber to control the chamber pressure. The ICP etching rate is determined to be ~7.5 nm/s associated with the conditions: the flow rate ratio $Cl_2$/Ar=10/25 with the ICP source power, bias power set at 200/200 and chamber pressure of 2.5 mTorr. The residual $SiO_2$ layer is subsequently removed by a buffer oxidation etchant (BOE).

The hybrid NQD/deep etched LED and the hybrid NQD/shallow etched LED are prepared by drop casting the colloidal NQDs onto the corresponding LEDs.

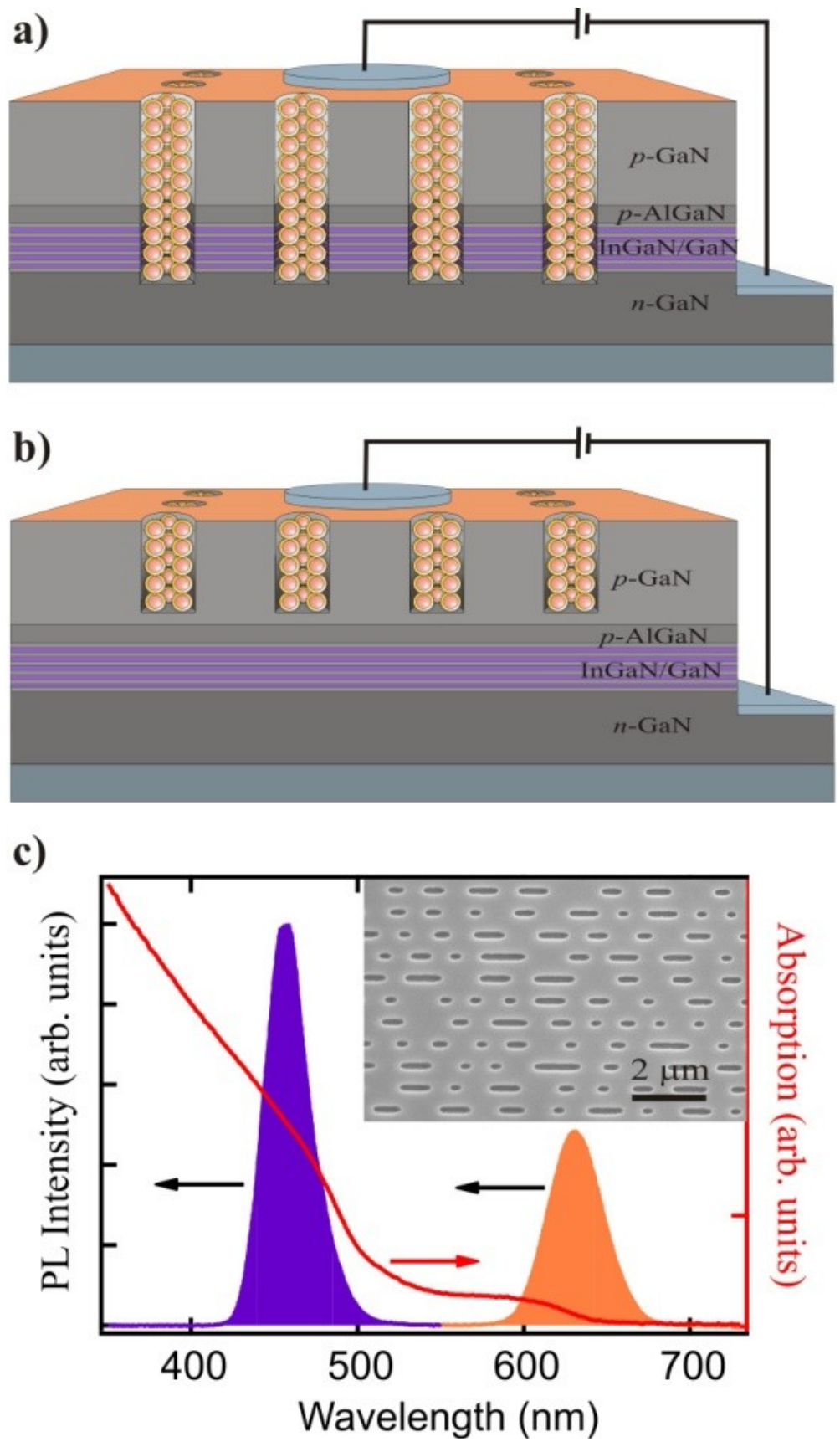

**Figure 1.** Schematic diagrams of a) the hybrid NQD/deep etched LED. b) the hybrid NQD/shallow etched LED. c) Spectral overlap of the QW emission (blue) and the NQD absorption (red line). NQD emission peak (orange) is around 630 nm at room temperature.

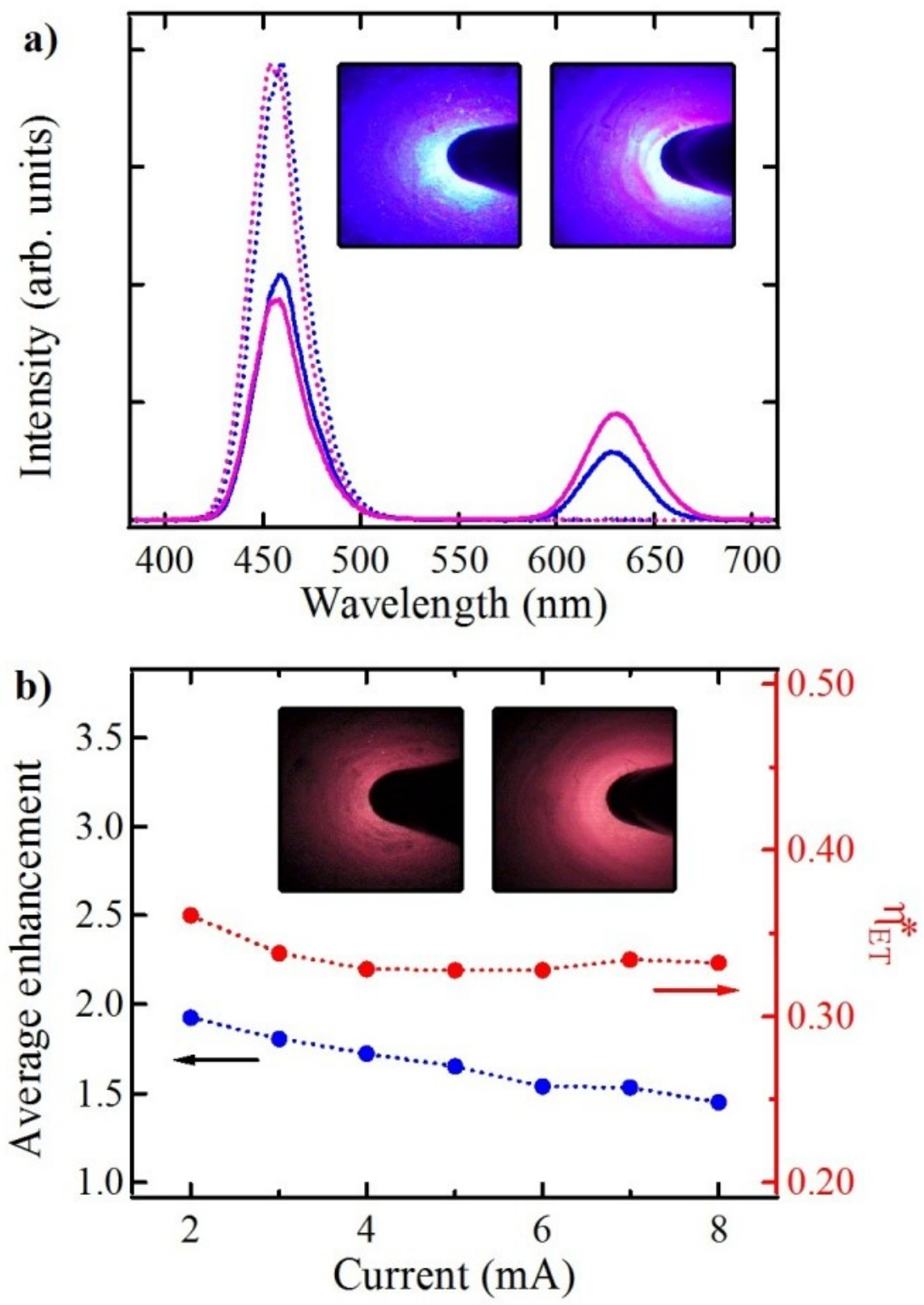

**Figure 2.** NQD emission enhancement and energy transfer efficiency of the hybrid color conversion light emitting diode. a) QW emission of the bare shallow etched LED (dotted blue line) and the bare deep etched LED (dotted pink line) at 3.8 mA and 7 mA respectively. The solid blue line (solid pink line) represents the corresponding EL of the hybrid NQD/shallow etched LED (hybrid NQD/deep etched LED). Inset: left (right) panel shows illumination of the hybrid shallow etched (deep etched) LED. b) Red solid circles represent effective nonradiative energy transfer efficiency at different injection current. Blue solid circles represent the enhancement of NQD emission. Inset: left (right) panel shows illumination of the NQDs in hybrid shallow etched (deep etched) LED.

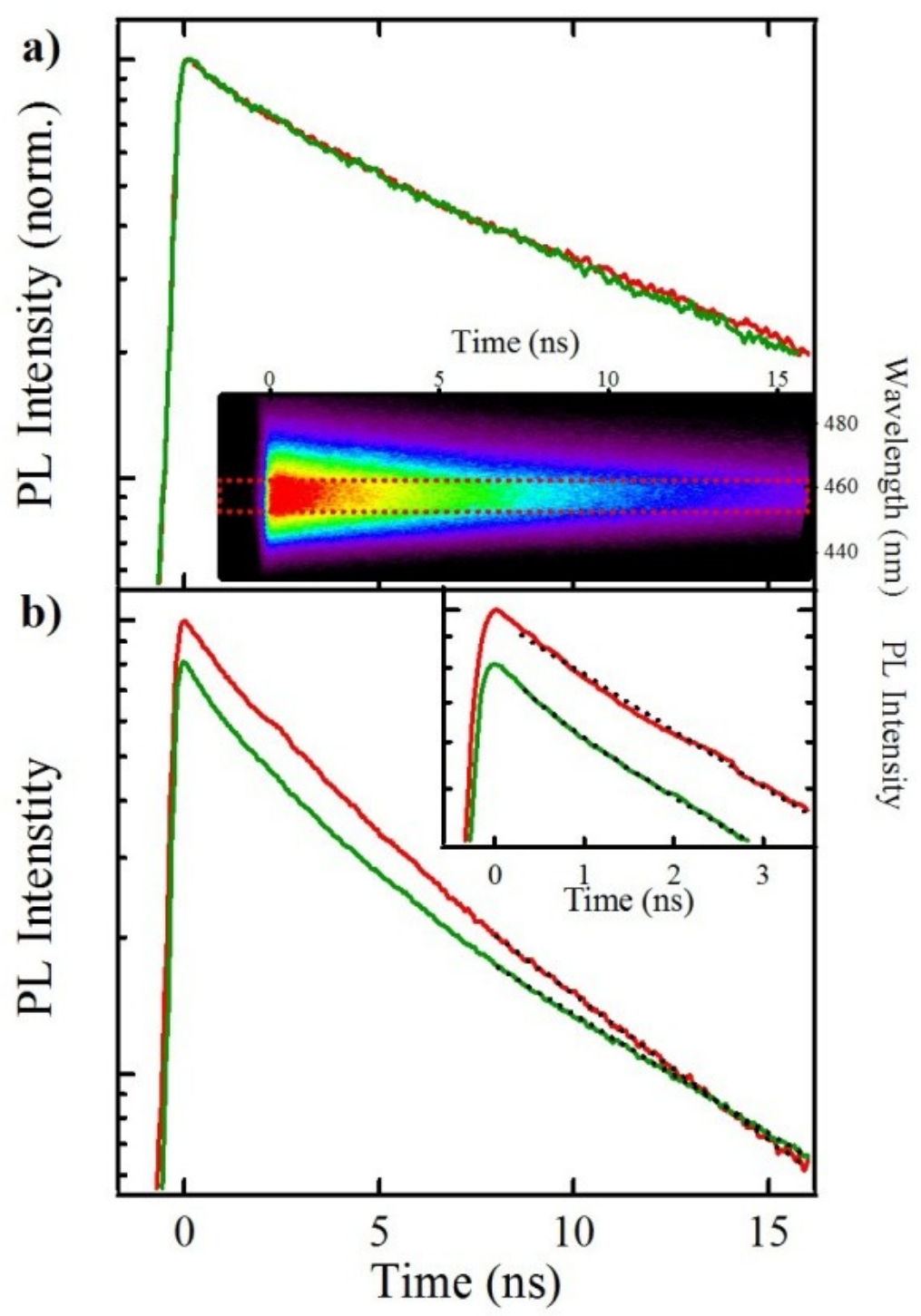

**Figure 3.** QW photoluminescence decay in the hybrid structures. a) QW emission decay of the shallow etched LED without (red solid line) and with (green solid line) NQDs. Inset shows the streak image of the bare shallow etched LED. b) QW emission decay of the deep etched LED without (red solid line) and with (green solid line) NQDs. Dotted lines are fittings described in text. Inset shows the effect of nonradiative energy transfer on the QW photoluminescence decay at early decay period.

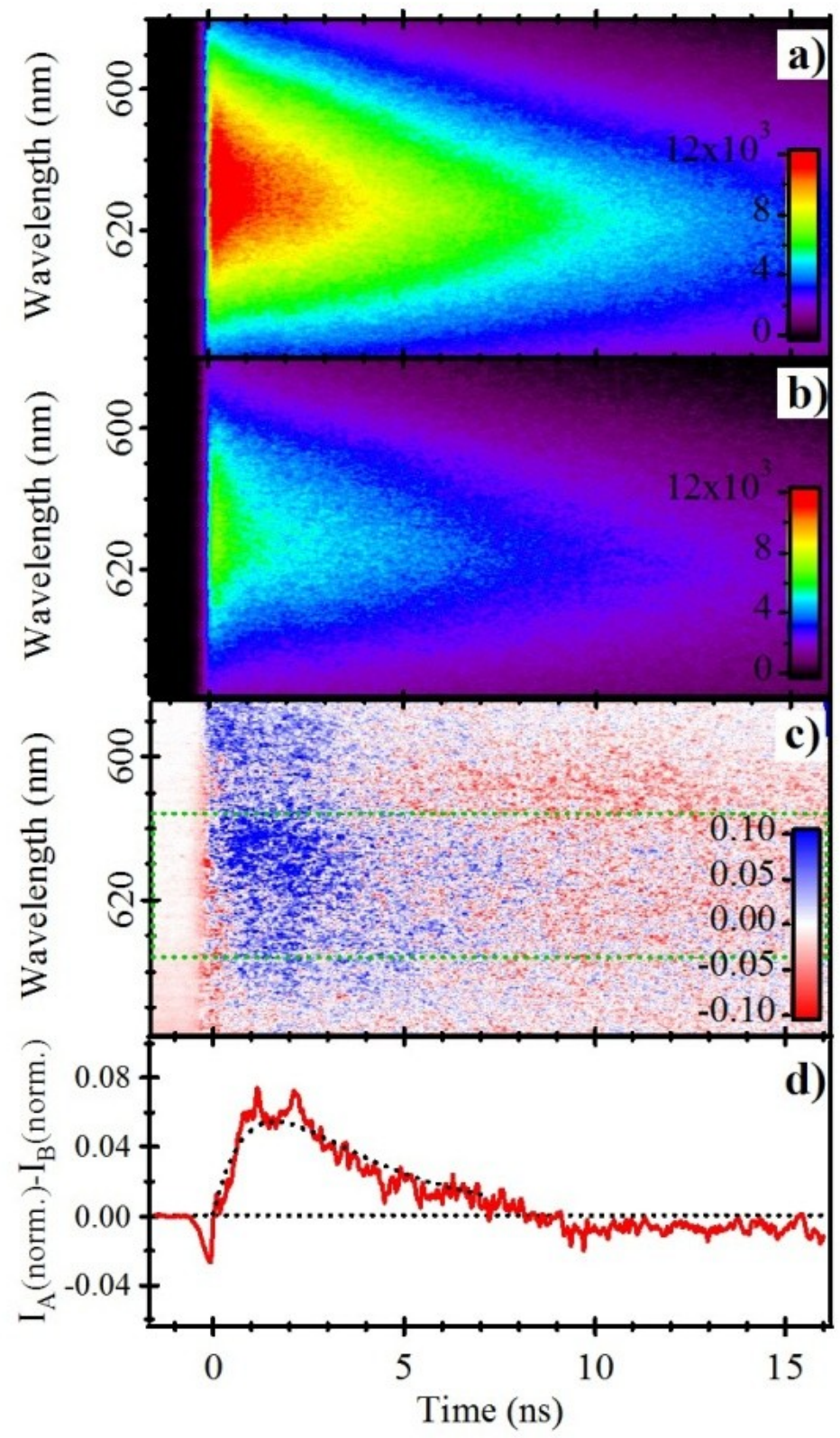

**Figure 4.** NQD fluorescence decay in the hybrid structures. a) NQDs on the deep etched LED. b) NQDs on the shallow etched LED. c) Difference of fluorescence dynamics calculated from the normalized data of a) and b). d) The average difference of fluorescence dynamics extracted from 20 nm energy window centered at the peak of NQD emission. Dotted line is the fitting described in text.